\documentclass[pre,aps,groupaddress,showpacs,prd,twocolumn]{revtex4}
\usepackage{epsfig,amsmath,graphicx,amssymb,overpic}
\usepackage{epstopdf}
\usepackage{colordvi}
\usepackage{color}
\def\be{\begin{equation}}
\def\ee{\end{equation}}
\def\bee{\begin{eqnarray}}
\def\ene{\end{eqnarray}}
\def\bes{\begin{subequations}}
\def\ees{\end{subequations}}
\newcommand{\br}{{\bf r}}
\newcommand{\bv}{{\bf v}}

\newcommand{\bc}{{\bf c}}
\newcommand{\ba}{{\bf a}}

\newcommand{\bb}{{\bf b}}

\begin{document}
\title{Three-dimensional rogue waves in non-stationary parabolic potentials}
\author{Zhenya Yan$^{1,2}$}
\email{zyyan_math@yahoo.com}
\author{V. V. Konotop$^3$}
\author{N. Akhmediev$^4$}
\affiliation{$^1$Key Laboratory of Mathematics Mechanization,
Institute of Systems Science, AMSS, Chinese Academy of Sciences,
Beijing 100080, China\\
$^2$International Centre for Materials Physics, Chinese Academy of
Sciences, Shenyang, 110016, China \\
$^3$Centro de F\'isica Te\'orica e Computacional and Departamento
de F\'isica, Faculdade de Ci\^enacias, Universidade de Lisboa,
Lisboa
1649-003, Portugal \\
$^4$Optical Sciences Group, Research School of Physics and
Engineering, Institute of Advanced Studies, The Australian
National University, Canberra ACT 0200, Australia}


\begin{abstract}

Using symmetry analysis we systematically present a
higher-dimensional similarity transformation reducing the
(3+1)-dimensional inhomogeneous nonlinear Schr\"odinger (NLS)
equation with variable coefficients and parabolic potential to the
(1+1)-dimensional NLS equation with constant coefficients. This
transformation allows us to relate certain class of localized exact
solutions of the (3+1)-dimensional case to the variety of solutions
of integrable NLS equation of (1+1)-dimensional case. As an example,
we illustrated our technique using two lowest order rational
solutions of the NLS equation as seeding functions to obtain rogue
wave-like solutions localized in three dimensions that have
complicated evolution in time including interactions between two
time-dependent rogue wave solutions. The obtained three-dimensional
rogue wave-like solutions may raise the possibility of relative
experiments and potential applications in nonlinear optics and BECs.

\end{abstract}

\maketitle


\section{Introduction}

Similarity analysis is one of modern powerful techniques which
allows us to find self-similar solutions of equations that
previously were known to be non-integrable (see, e.g., \cite{BK}
and references therein). They do not provide the complete
integrability. However, they help to produce selected solutions in
analytical form which may be important for variety of
applications. One of the representative examples is the nonlinear
Schr\"odinger (NLS) equation. It is well known that in
(1+1)-dimension ((1+1)-D) this equation is completely integrable
by inverse scattering technique~\cite{IST}. In (2+1)-D the
equation is not integrable. However, solutions localized in two
transverse directions do exist \cite{Chiao64} but may be unstable
and subjected to collapse \cite{Vlasov,Berge}. They are mostly
known from numerical simulations \cite{Akhmanov}. Remarkably, some
of the localized solutions can be found using similarity
reductions \cite{Gagnon}. Despite being unphysical, exact
solutions provide some insight on the properties of the equation
that is important for many applications. Clearly, adding a
dimension changes drastically integrability properties of the
equation. Thus, (3+1)-D NLS equation is not an exception and we
are faced with the problem of finding its solutions knowing that
they are not directly related to the solutions of the same
equation  in lower dimensionality.

The NLS equation in (3+1)-D  is an important model for variety of
physical problems~\cite{Optics, BEC}. It is used in nonlinear
optics~\cite{Optics}, condensed matter physics and in particular
in modelling Bose-Einstein condensate (BEC)~\cite{BEC}. Numerical
solutions can be found with various techniques but the value of an
analytical approach is significant by itself. In this work,
extending the ideas of ~\cite{yanvk,YH} we use the similarity
transformations to reduce the dimensionality of the equation from
(3+1)-D to (1+1)-D. In the former case, the coefficients in the
equation are variable while in the latter they can be chosen to be
constants. This allows us to use the complete integrability of the
(1+1)-D equation.

More specifically, we will focus on the possibility of constructing truly
three-dimensional rogue waves, i.e. waves whose dynamics essentially depends on all spatial coordinates,
 although it is possible to identify the coordinate in which the motion is effectively one-dimensional.
 This example is directly related to the description of matter wave dynamics in the mean-field
 approximation (where the NLS equation is also known as the Gross-Pitaevskii equation), thus representing a unique
 possibility of creating and observing three-dimensional rogue matter  waves.
We note that the conventional rogue waves are either
two-dimensional, as it happens, e.g., in the ocean~\cite{ocean}, in
wide aperture optical cavities~\cite{Arecchi} and in capillary wave
experiments~\cite{Shats} or one-dimensional and they appear in many
fields including nonlinear
 optics~\cite{Solli-Eggleton,Yeom,BKA-opt,ypla10}, cigar-shaped BECs~\cite{BKA}, atmosphere~\cite{at}, and
 finances~\cite{yan09}.

The rest of this paper is organized as follows. In Sec. II, we
describe the (1+1)-D similarity transformation reducing the (3+1)-D
inhomogeneous nonlinear Schr\"odinger (NLS) equation with variable
coefficients and parabolic potential to the (1+1)-D NLS equation
with constant coefficients. In Sec. III, we determine the
self-similar variables and constraints satisfied by the coefficients
in the (3+1)-D inhomogeneous NLS equation. Moreover, we give some
comments about these coefficients. Sec. IV mainly discusses two
types of localized 3D rogue wave-like solutions, which profiles are
exhibited. Finally, we give some conclusions in Sec. V.

\section{The 3D model and similarity reductions}

The original three-dimensional inhomogeneous NLS equation with
variable coefficients can be written in a dimensionless form: \bee
\label{nlsv} i\frac{\partial\Psi}{\partial
t}=-\frac{1}{2}\nabla^2\Psi+v(\br,t)\Psi+
g(t)|\Psi|^2\Psi+i\gamma(t)\Psi, \ene where the physical field
$\Psi\equiv\Psi(\br, t)$,\ $\br\in\mathbb{R}^3$, $\nabla
\equiv(\partial_x,\partial_y,\partial_z)$ with $\partial_x\equiv
\partial/\partial x$, the external potential $v(\br,t)$ is a real-valued
function of time and spatial coordinates, the nonlinear coefficient
$g(t)$ and gain/loss coefficient $\gamma(t)$ are real-valued
functions of time. This equation arises  in many fields such as
nonlinear optics (see, e.g.,~\cite{Optics}) and BECs (alias the
three-dimensional Gross-Pitaevskii equation with variable
coefficients, see, e.g., ~\cite{BEC, yanvk,YH}).

We search for a similar transformation connecting solutions of
Eq.~(\ref{nlsv}) with those of the (1+1)-D standard NLS equation
with constant coefficients, i.e.
 \bee \label{ODE}
\begin{array}{l}
 \displaystyle i\frac{\partial\Phi(\eta, \tau)}{\partial\tau}
 =-\frac{\partial^2\Phi(\eta, \tau)}{\partial \eta^2}+G|\Phi(\eta, \tau)|^2\Phi(\eta, \tau). \label{NLS}
\end{array} \ene
 Here the physical field $\Phi(\eta,\tau)$ is a function of
two variables $\eta\equiv \eta(\br,t)$ and $\tau\equiv \tau(t)$
which are  to be determined, and $G$ is a constant. Since our main
goal is to study three-dimensional rogue waves, we choose $G=-1$
which corresponds to the attractive case (or focusing nonlinearity
in optics and negative scattering lengths in the BEC theory). In
order to control boundary conditions at infinity we impose the
natural constraints~\cite{yanvk} \bee \label{constrain1} \eta\to
0\quad\mbox{at}\quad \br\to0\quad \mbox{and}\quad
\eta\to\infty\quad\mbox{at}\quad \br\to\infty. \ene

We are looking for the physical field $\Psi(\br, t)$ in the form
of the ansatz ~\cite{YH} \bee \label{tran}
\begin{array}{l}
\Psi(\br,t)=\rho(t)e^{i\varphi(\br,t)}\Phi[\eta(\br,t),\tau(t)]
\end{array}
\ene with $\rho(t)$ and $\varphi(\br,t)$ (like $\tau(t)$, and
$\eta(\br,t)$, introduced above) being the real-value functions of
the indicated variables, The ansatz~(\ref{tran})  allows us to
reduce the problem to (1+1)-D one (we notice that it  differs from
the one-dimensional stationary reductions~\cite{Juan2, yanvk}).
Variables in this reduction are to be determined from the
requirement for the new function $\Phi(\eta(\br,t),\tau(t))$ to
satisfy Eq.~(\ref{ODE}) (we notice that there also exist other
similar reductions for Eq.~(\ref{nlsv}) which require that
$\Phi(\eta,\tau)$ may satisfy other nonlinear equations). Thus, we
substitute the transformation (\ref{tran}) into Eq.~(\ref{nlsv})
and after relatively simple algebra obtain the system of partial
differential equations \bes \label{sys}\bee
 \label{sys1} && \nabla^2\eta=0, \ \  \eta_t+\nabla\varphi\cdot\nabla\eta=0,
 \ \   2\tau_t-|\nabla\eta|^2=0, \quad \\
 \label{sys4} &&  2\rho_t+\rho\nabla^2\varphi-2\gamma(t)\rho=0,\\
 \label{sys5} &&  2g(t)\rho^2-G|\nabla\eta|^2=0,  \\
 \label{sys6} &&  2v(\br,t)
 + |\nabla\varphi|^2+2\varphi_t=0.
\ene
\ees

Generally speaking, equations in the system (\ref{sys}) are not
compatible with each other when linear and nonlinear potentials
are arbitrary.
 One, however, can pose the problem to find the
functions
$v(\br,t)$, $g(t)$ and $\gamma(t)$ such that
the   system (\ref{sys}) becomes solvable.  This requirement leads us to the
procedure which can be outlined as follows.

\begin{itemize}
    \item Firstly, we solve Eq.~(\ref{sys1}) subject to the boundary
    conditions (\ref{constrain1}) thus obtaining
     the similarity variables $\eta(\br,t), \tau(t)$ and the phase $\varphi(\br,t)$.

    \item Secondly, we consider Eqs.~(\ref{sys4})-(\ref{sys6}) as
     definitions  for the functions $\rho(t), \ v(\br,t)$ and
    $g(t)$ in terms of already known functions $\eta(\br,t), \tau(t)$ and $\varphi(\br,t)$.
    \end{itemize}
Note that the first step determines transformation of variables
which does not involve explicitly any specific time dependent
coefficients. However, these coefficients may appear after
integration of these equations. The second step determines the
coefficients which are compatible with the above change of
variables. Thus, it leads to the model (\ref{NLS}). In this
approach, the function $\eta({\bf r},t)$ defines the surface where
the wave has a constant amplitude. The function $\varphi(\br,t)$
determines the wave-front solution (the manifold of the constant
phase).

Thus, we can establish a correspondence between selected solutions
of the (3+1)-dimensional inhomogeneous NLS equation with variable
coefficients (\ref{nlsv}) and known solutions of completely
integrable NLS equation (\ref{ODE}). The latter has an infinite
number of solutions thereby giving us a chance to look for
physically relevant solutions of the (3+1)-dimensional case. In
particular, we can relate them to the recently studied rogue wave
solutions of the NLS equation~\cite{AAS, AAT,ASA,ACA}. As a
consequence, we obtain three-dimensional time-dependant rogue wave
solutions of Eq.~(\ref{nlsv}).

\begin{figure*}[!ht]
\begin{center}
\vspace{0.01in}
{\scalebox{0.75}[0.7]{\includegraphics{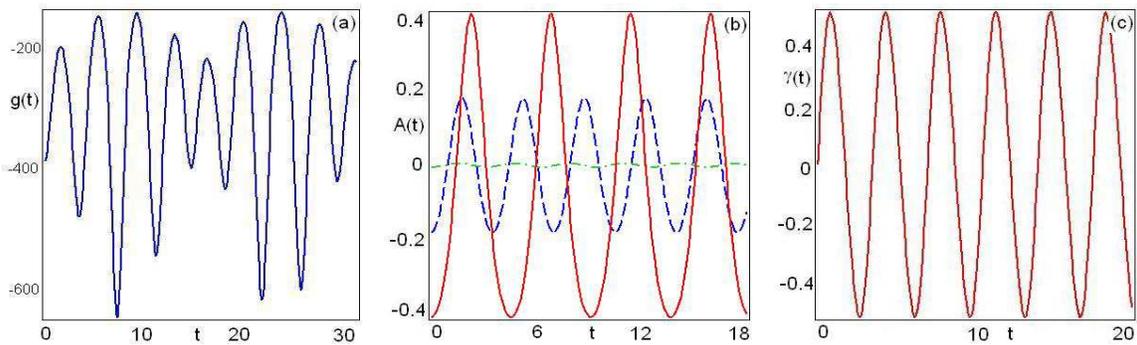}}}
\end{center}
\vspace{-0.25in} \caption{\small (color online). Profiles of (a)
nonlinearity $g(t)$ given by Eq.~(\ref{varb2}), (b) the coefficients
$A_x(t)$ (solid line), $A_y(t)$ (dashed line), and  $A_z(t)$
(dashed-dotted line) of second degree term of the linear potential
$v(\br,t)$ given by Eq.~(\ref{A}) and (c) the gain/loss term
$\gamma(t)$ vs time for the parameters are given by
Eq.~(\ref{para1}) with $k_x=0.9,\, k_y=0.6,\ k_z=0.1$ and $k=0.6$. }
\label{fig:0}
\end{figure*}

\section{Variables and coefficients of the transformation}

Solving Eq.~(\ref{sys1}) we can write the similarity variables
$\eta(\br,t),\ \tau(t)$ and the phase $\varphi(\br, t)$ in the
form
 \bes \label{var} \bee
 \label{var1} &&
 \eta(\br, t)=\bc(t)\cdot\br-\int^t_0\bc(s)\cdot\ba(s)ds, \\
\label{var2} &&
 \tau(t)=\displaystyle \frac12\int^t_{0}|\bc(s)|^2ds, \\
\label{var3} &&
\varphi(\br,t)=\br\,\hat\Omega(t)\,\br+\ba(t)\cdot\br+\omega(t)
 \ene \ees
 where we have introduced the diagonal
time-dependent 3$\times$3 matrix $\hat\Omega(t)={\rm
diag}(\Omega_x(t),\Omega_y(t),\Omega_z(t))$ with
$\Omega_{\sigma}(t)=-\dot{c}_{\sigma}(t)/[2c_{\sigma}(t)]$
(hereafter $\sigma=x, y, z$ and an overdot stands for the derivative with respect to time).
The coefficients $\bc (t)=(c_x(t), c_y(t), c_z(t))$, $\ba (t)=(a_x(t), a_y(t), a_z(t))$ and
 $\omega(t)$ are time-dependent functions.

Now, from Eqs.~(\ref{sys4})-(\ref{sys6}) we obtain the functions
$\rho(t), \ v(\br,t)$ and    $g(t)$ in the form
 \bes \label{varb}
\bee
 \label{varb1} &&
 \rho(t)=\rho_0\sqrt{|c_x(t)\,c_y(t)\,c_z(t)|}\,e^{\int^t_0\gamma(s)ds}, \\
 \label{varb2} &&
 g(t)=\frac{G\,|\bc(t)|^2}{2\rho_0^2\,|c_x(t)\,c_y(t)\,c_z(t)|\,e^{2\int^t_0\gamma(s)ds}}, \\
  &&
  v(\br,t) =\br\,\hat A(t)\,\br+\bb (t)\cdot \br -\dot\omega(t) -\frac12
  |\ba(t)|^2,
\label{varb3}
\ene  \ees
 where $\rho_0$ is an integration constant.

A few comments would be useful here. First, the gain/loss term
$\gamma(t)$, is determined in the initial statement of the problem
and can serve as an additional control function or a parameter if it
is a constant. Then changing the time-dependent dissipation we can
excite different dynamical regimes. Second, the change of the all
parameters is interrelated. In practical terms such a time
dependence can be performed in different ways for different physical
systems. In particular, in the context of the BEC applications, this
can be done by simultaneous change of the frequency of the lasers
controlling the external trap $v(\br, t)$ and the detuning from the
Feshbach resonance, responsible for the variation of $g(t)$.
Finally, we notice that one can consider the case of
$g(t)\equiv$const which however is reduced to the trivial
 case of a plane wave, whose parameters change along the chosen direction (determined by the vector $\bc$ which is a
 constant in this case). Such solutions will not be considered here.

In writing the linear potential $v(\br,t)$ we have defined the
diagonal time-dependent 3$\times$3 matrix $\hat
A(t)=$diag$(A_x(t), A_y(t), A_z(t))$ with the entries \bee
\label{A}
 A_{\sigma}(t)
 =\frac{\ddot{c}_{\sigma}(t)}{2c_{\sigma}(t)}-\frac{\dot{c}_{\sigma}^2(t)}{c_{\sigma}^2(t)}
    \ene
 and the vector function $\bb(t)=(b_x(t), b_y(t), b_z(t))$ with the entries
 \bee \label{B}
  b_{\sigma}(t)=\frac{\dot{c}_{\sigma}(t)\,a_{\sigma}(t)}{c_{\sigma}(t)}-\dot{a}_{\sigma}(t).
\ene

 It is easy to see that the velocity field
$\bv(\br,t)=\nabla\varphi(\br,t)$ corresponding to the
above-mentioned phase $\varphi(\br,t)$ is given by \bee
 \begin{array}{l}
 \bv(\br,t)=2(\Omega_x(t)x, \,  \Omega_y(t)y, \,
 \Omega_z(t)z)+\ba(t) \quad
 \end{array}
\ene such that we have the divergence of the vector field
$\bv(\br,t)$ in the form \bee
 \begin{array}{rl}
 {\rm div}\, \bv (\br,t) &
   =2[\Omega_x(t)+\Omega_y(t)+\Omega_z(t)] \vspace{0.08in}\cr
   &=-\partial_t\ln|c_x(t)c_y(t)c_z(t)|. \end{array}
   \ene
   Thus the zeros of any of the components of $\bc(t)$ means the divergence of the field, which
    occurs  at the instants when the nonlinearity $g(t)$ becomes infinite
    [see (\ref{varb2})]. Such cases will not be considered  in the present paper.

It follows from Eqs.~(\ref{varb3}) and (\ref{A}) that if we
require that the linear potential $v(\br,t)$ is a second degree
polynomial for every space $x,\ y,\ z$, then we have
$A_{\sigma}\not=0$, i.e.,
$c_{\sigma}\ddot{c}_{\sigma}-2\dot{c}_{\sigma}^2\not=0$, which
denote that $c_{\sigma}$ are not equivalent to constants but some
functions of time. These time-dependent functions $c_{\sigma}(t)$
will affect on the other variables (see
Eqs.~(\ref{var1})-{\ref{B})) such that self-similar solutions of
Eq.~(\ref{nlsv}) in the form (\ref{tran}) exhibit abundant
structures. In what follows we will use specific solutions (e.g.,
rogue wave solutions) of the NLS equation to illustrate the
nontrivial dynamics of three-dimensional rogue wave-like solutions
defined by the Eq.~(\ref{nlsv}) for the different parameters
mentioned above.

\section{Two types of localized 3D rogue wave-like solutions}

As two representative examples, we consider the lowest order
rational solutions of the NLS equation which serve as prototypes of
rogue waves. First, we use the first order rational solution of
Eq.~(\ref{ODE}) (see \cite{AAS}). As a result, we obtain the
first-order non-stationary rogue wave solutions of Eq.~(\ref{nlsv})
in the form
 \bee
\nonumber \Psi_1(\br,t)=\rho_0\sqrt{|c_x(t)\,c_y(t)\,c_z(t)|}\,e^{\int^t_0\gamma(s)ds} \qquad\qquad \\
\times
   \left[1-\frac{4+8i\tau(t)}{1+2\eta^2(\br,t)+4\tau^2(t)}\right]e^{i[\varphi(\br,t)+\tau(t)]},
\label{solu1} \ene where the variables $\eta(\br,t),\ \tau(t)$ and
the phase $\varphi(\br,t)$ are given by
Eqs.~(\ref{var1})-(\ref{var3}).

For the illustrative purposes, we can choose these free parameters
in the form
\bee \label{para1}
\begin{array}{l}
 c_{\sigma}(t)=a_{\sigma}(t)={\rm dn}(t, k_{\sigma}),
\vspace{0.05in}\cr \rho_0=1,\,\gamma(t)={\rm sn}(t,k){\rm
cn}(t,k),
\end{array}
\ene (where dn, sn, and cn stand for the respective Jacobi
elliptic functions, and $k_{\sigma}, k$ are their moduli.) and
$\omega(t)=0$. Figure \ref{fig:0} depicts the profiles of
nonlinearity
 $g(t)$ given by Eq.~(\ref{varb2}), the coefficients of second degree terms of the linear potential
 $v(\br,t)$ given  by Eq.~(\ref{A}) and
 the gain/loss term $\gamma(t)$ vs time for the chosen parameters given by Eq.~(\ref{para1}). The
evolution of intensity distribution of the 3D field (\ref{solu1})
is shown in Fig.~\ref{fig:1}.
 We can see that the simple Lorenzian
 function of the (1+1)D case is transformed into a significantly more complicated evolution along the
 $t$-axis. The solution is localized in space and keeps the localization infinitely in
 time, which differs from the usual rogue wave solutions (see \cite{ASA2}).

\begin{figure*}
\begin{center}
\vspace{0.01in}
{\scalebox{0.76}[0.76]{\includegraphics{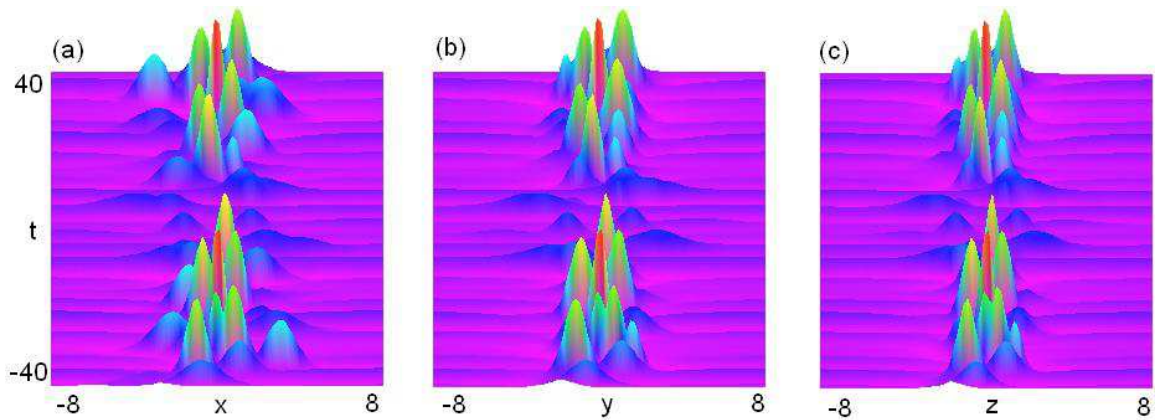}}}
\end{center}
\vspace{-0.25in} \caption{\small (color online). Color coded plot
of wave intensity (a) $|\Psi_1|^2(x,0,0,t)$ with
$\max_{\{x,0,0,t\}}|\Psi_1|^2=0.06$, (b) $|\Psi_1|^2(0,y,0,t)$
with $\max_{\{0,y,0,t\}}|\Psi_1|^2=0.068$ and (c)
$|\Psi_1|^2(0,0,z,t)$ with $\max_{\{0,0,z,t\}}|\Psi_1|^2=0.068$
defined by the solution (\ref{solu1}) for the parameters given by
Eq.~(\ref{para1}) with $k_x=0.9, \, k_y=0.6,\, k_z=0.1,\ k=0.6$.
Note that for given set of parameters, the wave is localized in
space.} \label{fig:1}
\end{figure*}

 On the other hand, if we choose the free parameters in the form
 \bee \label{para2}
\begin{array}{l}
 c_x(t)=a_x(t)=1+c_0\sin(t),  \cr c_y(t)=a_y(t)=1.2+c_0\cos(t), \cr
  c_z(t)=a_z(t)=0.8+c_0\sin(t), \   c_0=0.01
\end{array} \ene
and $\rho_0,\, \omega(t)$ and $\gamma(t)$ are same as the ones
given by Eq.~(\ref{para1}), then the evolution of intensity
distribution of the 3D rogue wave solutions (\ref{solu1}) will be
changed. Figure \ref{fig:20} displays the profiles of nonlinearity
 $g(t)$ given by Eq.~(\ref{varb2}) and the coefficients of second degree terms of the linear potential
 $v(\br,t)$ given  by Eq.~(\ref{A})  {\it vs} time for the chosen parameters given by Eq.~(\ref{para2}).
For this case, the 3D rogue wave solution (\ref{solu1}) is shown
 in Figs.~\ref{fig:4}. The solution is localized both in time and in space thus revealing
 the usual ``rogue wave" features. It is worth emphasizing here, that although the ``generating"
 function $\bc(t)$ was chosen as monochromatic function, the respective change of the nonlinearity $g(t)$ required for
 the existence of the exact solution is periodic but depending on various frequencies. This is natural
 reflection of the fact that we are dealing with a nonstationary solution of the nonlinear problem,
 characterized by the generation of multiple frequencies.

Generally speaking, we have large degree of freedom in choosing the coefficients of transformation.
As a result, we can describe infinitely large class of solutions of three-dimensional NLS equation
with every exact solution of the one-dimensional NLS equation. Additional possibility of choosing
the solution of the latter one increases tremendously variety of solutions that we can obtain.

\begin{figure}[!ht]
\begin{center}
\vspace{0.01in}
{\scalebox{0.4}[0.38]{\includegraphics{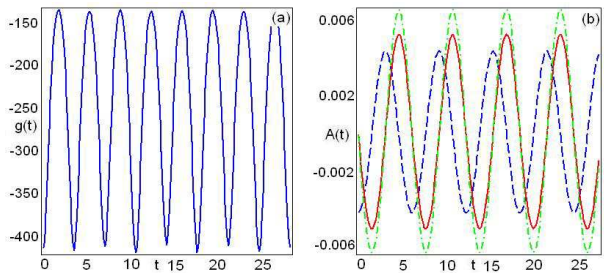}}}
\end{center}
\vspace{-0.25in} \caption{\small (color online). Profiles of (a)
nonlinearity $g(t)$ given by Eq.~(\ref{varb2}) and (b) the
coefficients $A_x(t)$ (solid line), $A_y(t)$ (dashed line), and
$A_z(t)$ (dashed-dotted line) of second degree term of the linear
potential $v(\br,t)$ given by Eq.~(\ref{A}) vs time for the
parameters are given by Eq.~(\ref{para2}).} \label{fig:20}
\end{figure}

\begin{figure*}[!ht]
\begin{center}
\vspace{0.01in}
{\scalebox{0.76}[0.8]{\includegraphics{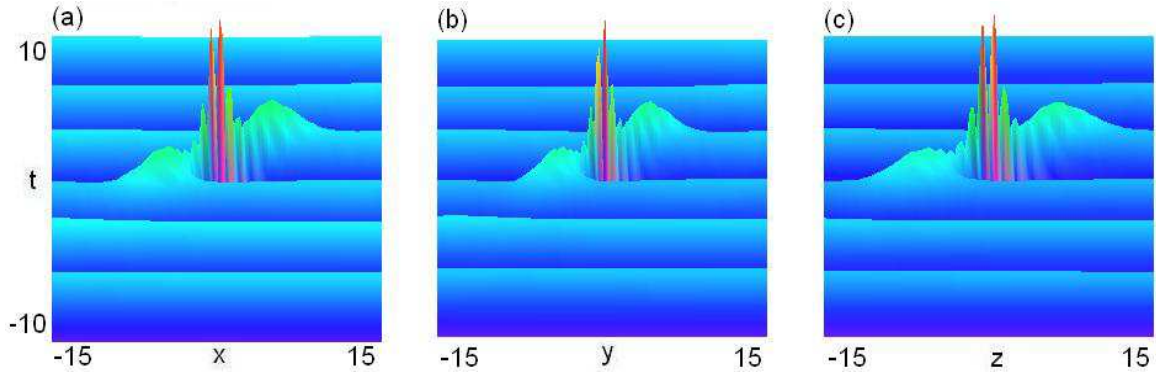}}}
\end{center}
\vspace{-0.25in} \caption{\small (color online). Color coded plot
of wave intensity (a) $|\Psi_1|^2(x,0,0,t)$ with
$\max_{\{x,0,0,t\}}|\Psi_1|^2=0.027$, (b) $|\Psi_1|^2(0,y,0,t)$
with $\max_{\{0,y,0,t\}}|\Psi_1|^2=0.03$ and (c)
$|\Psi_1|^2(0,0,z,t)$ with $\max_{\{0,0,z,t\}}|\Psi_1|^2=0.029$
defined by the solution (\ref{solu1})
 for the parameters given by Eq.~(\ref{para2}).} \label{fig:4}
\end{figure*}

When a higher order rational solution of the NLS equation
(\ref{ODE}) (see \cite{AAS}) is applied to the transformation
(\ref{tran}), we obtain the second-order non-stationary rogue wave
solutions of Eq.~(\ref{nlsv}) in the form \bee
\nonumber \Psi_2(\br,t)=\rho_0\sqrt{|c_x(t)\,c_y(t)\,c_z(t)|}\,e^{\int^t_0\gamma(s)ds} \qquad\qquad \\
\times
\left[1+\frac{P(\eta,\tau)-i\tau(t)\,Q(\eta,\tau)}{H(\eta,\tau)}\right]e^{i[\varphi(\br,t)+
\tau(t)]}, \label{solu2} \ene where these functions $P(\eta,\tau),
\ Q(\eta,\tau)$ and $H(\eta,\tau)$ are given by \cite{AAT}
 \bee \label{solu20}
 \begin{array}{l}
 P(\eta,\tau)=\displaystyle -\frac{1}{2}\eta^4-6\eta^2\tau^2-10\tau^4-\frac{3}{2}\eta^2
 -9\tau^2+\frac38,
\vspace{0.1in}\cr Q(\eta,\tau)=\displaystyle
\eta^4+4\eta^2\tau^2+4\tau^4-3\eta^2+2\tau^2-\frac{15}{4},
\vspace{0.1in}\cr
 H(\eta,\tau)=\displaystyle
 \frac{1}{12}\eta^6+\frac12\eta^4\tau^2+\eta^2\tau^4+\frac23\tau^6
  +\frac18\eta^4 \vspace{0.1in}\cr
  \displaystyle \qquad\qquad\quad
 +\frac92\tau^4 -\frac32\eta^2\tau^2+\frac{9}{16}\eta^2+\frac{33}{8}\tau^2+\frac{3}{32}.\end{array} \ene
 The variables $\eta(\br,t),\ \tau(t)$ and the phase $\varphi(\br,t)$ here are given by Eqs.
(\ref{var1})-(\ref{var3}).

As in the previous cases, we choose the parameters given by
Eqs.~(\ref{para1}) and (\ref{para2}) except for $a_{\sigma}(t)=0$.
The intensity distributions of the second-order rogue wave
solutions (\ref{solu2}) are depicted in Figs.~\ref{fig:2} and
~\ref{fig:5}. Clearly, the field evolution in this case is more
complicated. In one case, shown in Fig.~\ref{fig:2}, the solution
is localized in all three dimensions in space. In the other case,
shown in Fig.~\ref{fig:5} the solution is localized both in space
and in time thus displaying the basic feature of a rogue wave that
``appears from nowhere and disappears without a trace".

It follows from the above-mentioned two cases for the parameters
that the parameters $c_{\sigma}(t),\ a_{\sigma}(t)$ and
$\gamma(t)$ can be used to control the wave propagations related
 to rogue waves, which may raise the
possibility of relative experiments and potential applications in
nonlinear optics and BECs. Similarly we can also obtain
three-dimensional higher-order time-dependent rogue wave solutions
of Eq.~(\ref{nlsv}) in terms of the transformation (\ref{tran})
and higher-order rogue wave solutions of the NLS equation
(\ref{ODE}), which are omitted here.

As always happen with the nonlinear Sch\"odinger equation in two
and three dimensions, their localized solutions may collapse.
Stability of the solutions presented in our work is still an open
question. This question deserves separate studies  as it is a task
that is far from being trivial. We leave these studies to later
publications.

\begin{figure*}[!ht]
\begin{center}
\vspace{0.2in}
{\scalebox{0.76}[0.76]{\includegraphics{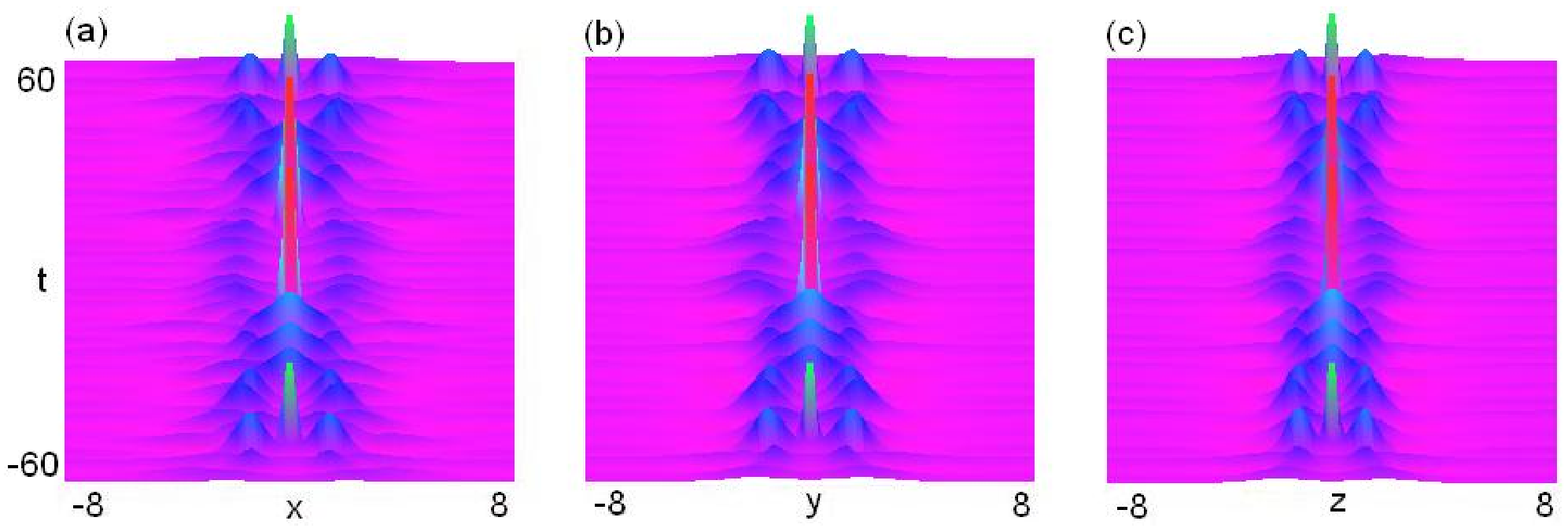}}}
\end{center}
\vspace{-0.25in} \caption{\small (color online). Color coded wave
intensity (a) $|\Psi_2|^2(x,0,0,t)$ with
$\max_{\{x,0,0,t\}}|\Psi_2|^2=0.135$, (b) $|\Psi_2|^2(0,y,0,t)$
with $\max_{\{0,y,0,t\}}|\Psi_2|^2=0.13$ and (c)
$|\Psi_2|^2(0,0,z,t)$ with $\max_{\{0,0,z,t\}}|\Psi_2|^2=0.125$
given by the solution (\ref{solu2}) for $a_{\sigma}(t)=0$ and
other parameters given by Eq.~(\ref{para1}) with $k_x=0.9, \,
k_y=0.6,\, k_z=0.1$ and $k=0.6$.} \label{fig:2}
\end{figure*}

\begin{figure*}[!ht]
\begin{center}
\vspace{0.01in}
{\scalebox{0.76}[0.8]{\includegraphics{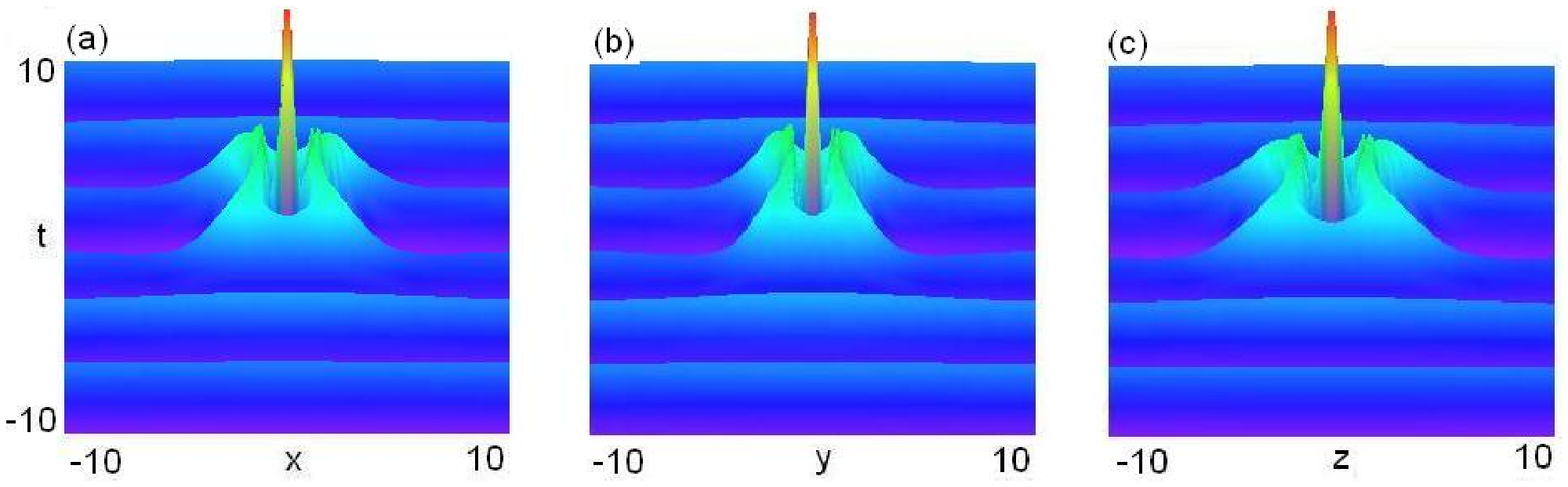}}}
\end{center}
\vspace{-0.25in} \caption{\small (color online).  Color coded wave
intensity (a) $|\Psi_2|^2(x,0,0,t)$ with
$\max_{\{x,0,0,t\}}|\Psi_2|^2=0.038$, (b) $|\Psi_2|^2(0,y,0,t)$
with $\max_{\{0,y,0,t\}}|\Psi_2|^2=0.036$ and (c)
$|\Psi_2|^2(0,0,z,t)$ with $\max_{\{0,0,z,t\}}|\Psi_2|^2=0.038$
given by the solution (\ref{solu2}) for $a_{\sigma}(t)=0$ and
other parameters given by Eq.~(\ref{para2}).}
\label{fig:5}\end{figure*}

\section{Conclusions}

In conclusion, we have presented similarity reductions of the
(3+1)-dimensional  inhomogeneous nonlinear Schr\"odinger equation
with variable coefficients to (1+1)-dimensional one with constant
coefficients. This transformation allows us to relate certain class
of localized solutions of the (3+1)-dimensional case to the variety
of solutions of integrable NLS equation of (1+1)-dimensional case.
As an example, we illustrated our technique by two lowest order
rational solutions of the NLSE. These are transformed into rogue
wave solutions localized in 3D space that have complicated evolution in
time. The technique may also be extended to other NLS-type equations
to exhibit their rogue wave solutions.

\acknowledgments ZYY gratefully acknowledges the support of the
NSFC60821002/F02. The research of VVK was partially supported by the
grant PIIF-GA-2009-236099 (NOMATOS) within the 7th European
Community Framework Programme. NA gratefully acknowledges the
support of the Australian Research Council (Discovery Project number
DP0985394).


\end{document}